\documentclass[12pt]{elsarticle}

\usepackage{graphicx}

\newcommand{\be}{\begin{equation}}
\newcommand{\ee}{\end{equation}}
\newcommand{\bea}{\begin{eqnarray}}
\newcommand{\eea}{\end{eqnarray}}
\newcommand{\ba}{\begin{eqnarray}}
\newcommand{\ea}{\end{eqnarray}}
\newcommand{\Dslash}{D\hspace{-1.6ex}/\hspace{0.6ex} }

\newcommand{\snchi}{\chi\hspace{-0.8ex}/\hspace{0.6ex} }
\begin{document}

\begin{frontmatter}


\title{On Chiral Symmetry Breaking, \\  Topology and Confinement  }

\author{Edward Shuryak}

\address{Department of Physics and Astronomy, \\ Stony Brook University,
Stony Brook, NY 11794 USA}
\begin{abstract} We start with the relation between the chiral symmetry breaking and gauge field topology. New lattice result further
enhance the notion of Zero Mode Zone, a very narrow strip of states with quasizero Dirac eigenvalues. Then we move to the issue
of ``origin of mass" and Brown-RHo scaling: a number of empirical facts contradicts to the idea that masses of quarks
and such hadrons as $\rho,N$ decrease near $T_c$. 
 We argue that while at $T=0$ the main contribution to the effective quark mass is chirally odd $m_{\snchi}$,
near $T_c$   it rotates to chirally-even component $m_\chi$, because ``infinite clusters" of topological solitons gets split into finite ones. 
Recent progress in understanding of topology require introduction of nonzero holonomy $<A_0>\neq 0$, which splits instantons into $N_c$
(anti)selfdual ``instanton-dyons". Qualitative progress, as well as first numerical studios of the dyon ensemble are reported.
New connections between chiral symmetry breaking and confinement  are recently understood, since instanton-dyons generates 
  holonomy potential with a minimum at confining value, if the ensemble is dense enough. 
\end{abstract}
 \end{frontmatter}


\section{Introduction}

Like other authors of this volume, I am much indebted to Gerry Brown. His decision to make me  
 his successor,  as a leader of  Stony Brook 
Nuclear Theory, was obviously the highest honor of my life. 
Twenty years of nearly daily  discussion with Gerry  about physics, life and life in physics
have not resulted in many common papers. Yet those thousands of hours were invaluable, especially for me,
after another twenty plus years in a relative solitude in Novosibirsk. 
Layers upon layers of knowledge came from Gerry, on science, scientists and life, with good share of his characteristic jokes.

  It was not easy to select the topic for this article.
Gerry was seriously excited a decade ago, while --induced by strongly coupled QGP 
-- Ismail Zahed and myself returned to the fate of the Coulomb  
bound states at the coupling approaching or exceeding the critical value:
a subject close to his heart from large-Z atoms and Birmingham days. While progress since then include strongly coupled quarkonia in
AdS/CFT  and  observations of supercritical resonances in graphene, 
 theoretically this problem  remains  basically unsolved.
On the other hand,  progress in fields I was mostly involved lately -- hydrodynamical description of
 higher flow harmonics in heavy ion collisions,  or of the ``explosive" high multiplicity
$pA$ and even $pp$  --
would not be so exciting to Gerry. So
  I decided to return to the core issues of our science -- chiral symmetry breaking, confinement and gauge topology. Slow but  steady progress
  is there, not much known
outside of a narrow circle. It would interest Gerry for sure.  

\begin{figure}[!b]\vskip 0.02in
 \center{ \hskip 0in\includegraphics[width=4cm]{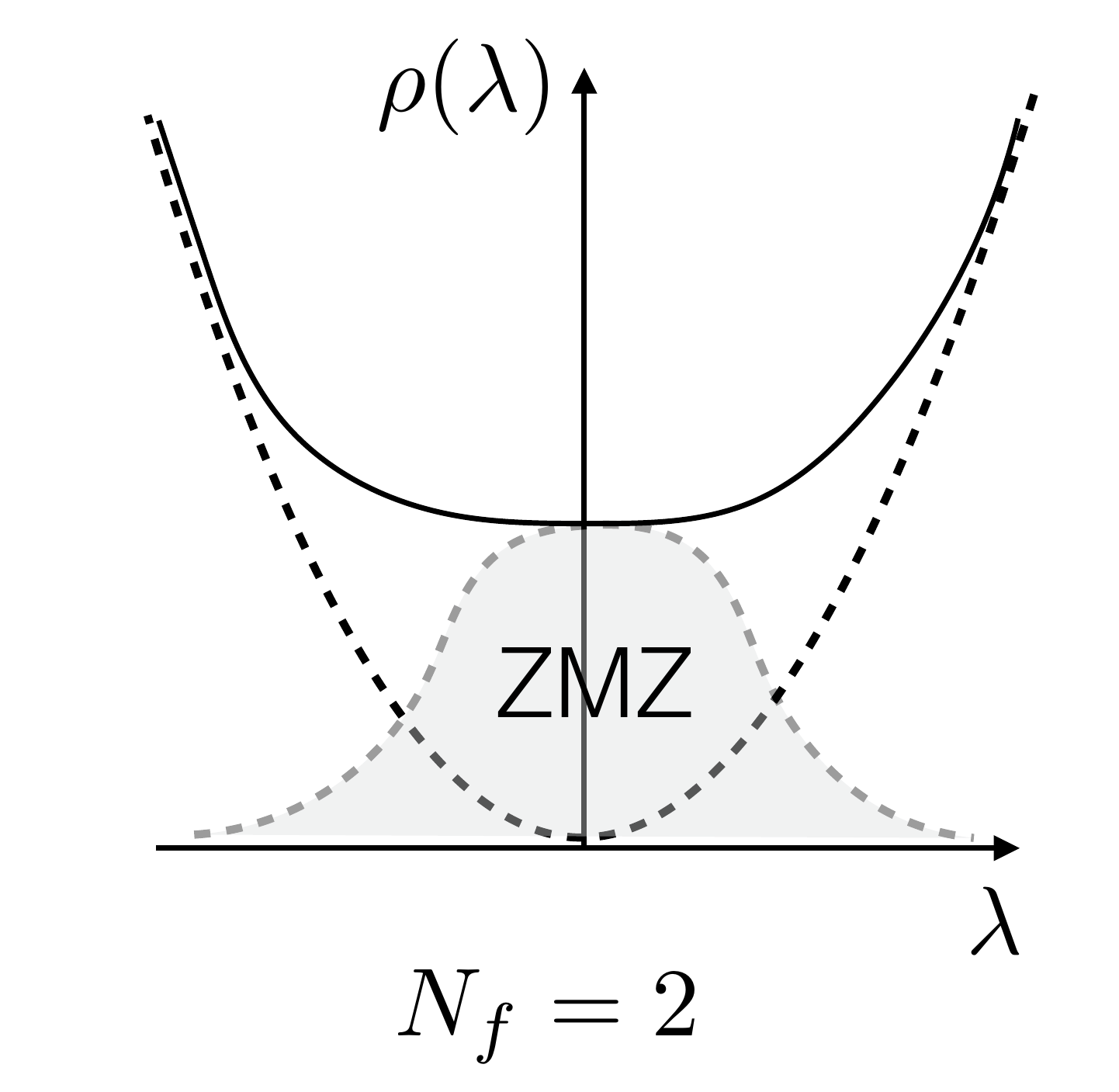}
 \includegraphics[width=3.7cm]{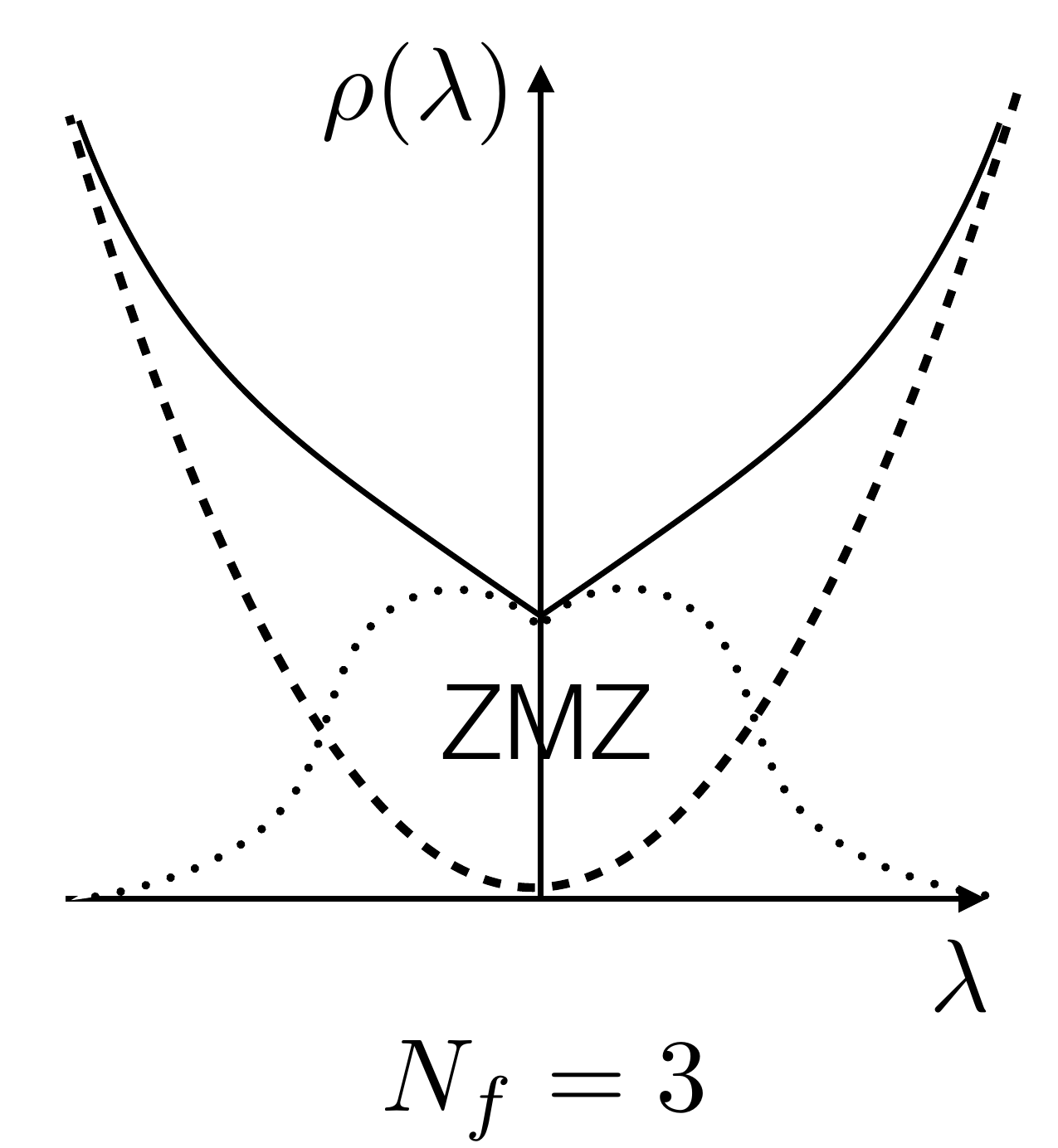}
 \includegraphics[width=4cm]{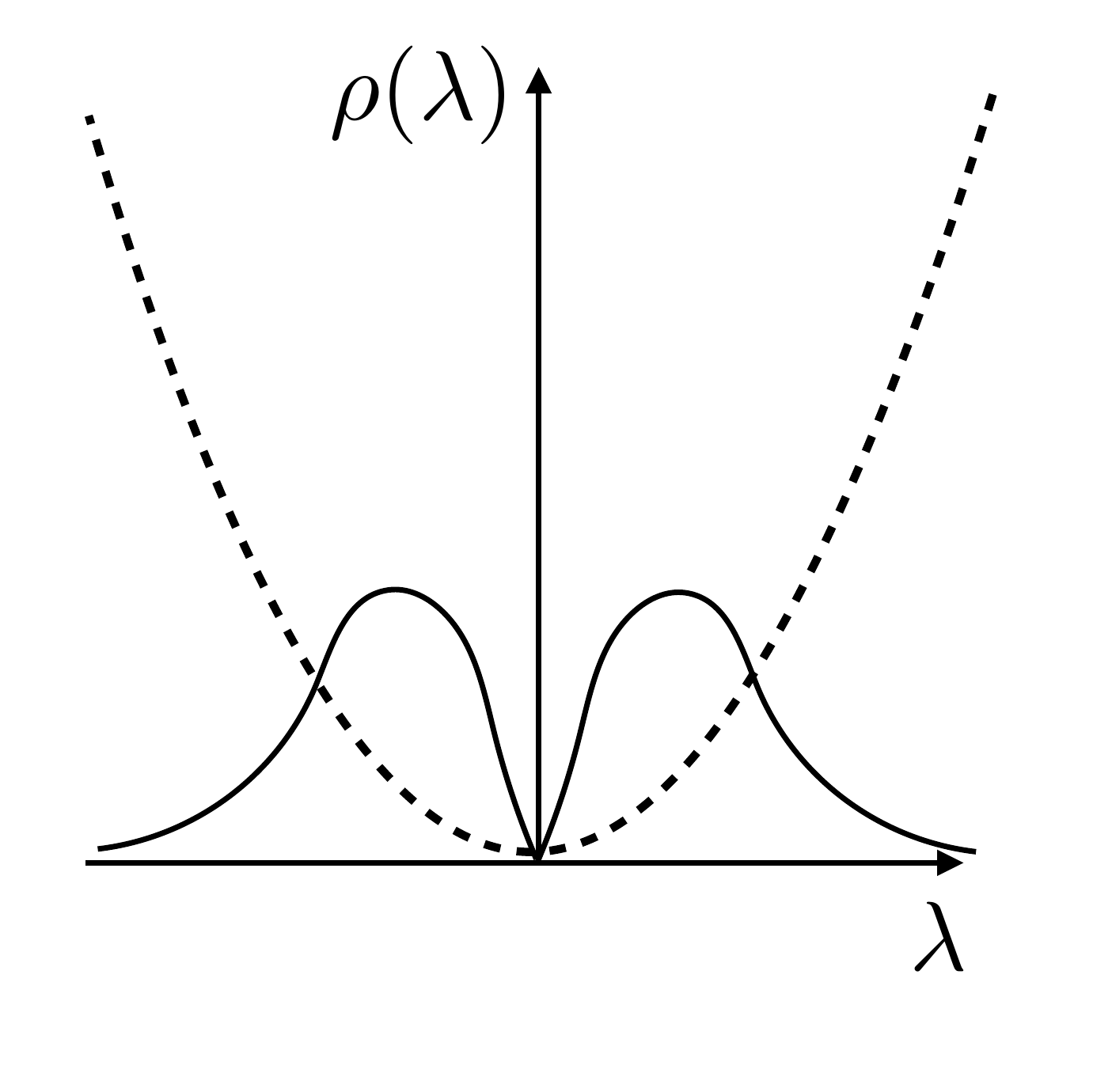}
 }
 \caption{\label{fig_ZMZ} The density of Dirac eigenvalues for $N_f=2$ (left), $N_f=3$ (middle) and in the critical case of restoring chiral symmetry (right) }
 \end{figure}
 
\section{The  Chiral Symmetry Breaking and the Zero Mode Zone  }
One way to describe this phenomenon --  a "3-d manybody" one -- goes back to Nambu-Jona-Lasinio (NJL) paper based on 
 an analogy to the BCS theory of superconductivity. A 
4-fermion attraction at soft momenta $|k|<\Lambda$, if strong enough, leads to a nonzero quark condensate 
ant a gap, at the surface of the Dirac sea.

Another approach --  a "4-dimensional single body" one -- is simpler to explain and to work with, in Euclidean setting.  Dirac
eigenvalues can be defined for any gauge fields configuration
$ \Dslash \psi_\lambda =\lambda \psi_\lambda $
and those may have finite or zero density of states $\rho(\lambda)$ at $\lambda \rightarrow 0$.  The former case breaks the chiral symmetry, and the condensate is just proportional to  $\rho(\lambda=0)\neq 0$ \cite{Banks:1979yr}. The alternative case $\rho(0)= 0$ is  chirally symmetric.
 This is reminiscent to the density of states  at the Fermi surface: if nonzero it definds a conductivity and many other   properties of a {\em conductor}, if zero it makes it an insulator.
 In Fig.\ref{fig_ZMZ}  we sketch the shapes of such density of states\footnote{Note that for brevity we do not discuss finite-size effects, assuming macroscopic limit $V\rightarrow \infty$ is already taken. Note also that
 for
 $N_f>2$ there exists the so called {\em Stern-Smilga cusp} $(\rho(\lambda)-\rho(0))\sim (N_f^2-4) |\lambda|$ in the middle.}
 , for  $N_f=2$ (left) and $N_f=3$ (middle).    The right picture
 corresponds to the critical case, when $\rho(0)$ vanishes and the chiral symmetry is being restored. At $T>T_c$ there appears a finite gap around $\lambda=0$,
 like in an insulator. 
 
   The idea that only a tiny subset of states near the Fermi surface dominates the physics is one of the pillars of 20-th century condensed matter theory.  Similar fundamental
 concept  is the {\em Zero Mode Zone} (ZMZ) introduced in 
  the context of the instanton liquid model (ILM) \cite{Shuryak:1981ff}.  The topological index theorems 
demand a connection between  the topological charge of the gauge fields and the number of {\em exactly zero}
eigenvalues.  Thus a collection of well-separated instantons and antiinstantons produce many  $\lambda=0$ states, if interactions are neglected.
If those are included, those states get {\em collectivised}, creating ZMZ, of various shapes we showed above. Details about
numerical simulations 
of instanton ensembles and actual ZMZ can be found in a review \cite{Schafer:1996wv}. Lattice studies confirmed those, and show that the ZMZ
states mix little with the ``plane wave"-type states of the perturbation theory.

   One crucial prediction of the ILM is the {\em surprisingly small width} of the ZMZ. 
In the instanton liquid model an amplitude describing ``hopping"
of a quark from one instanton to the next determines this width,   parametrically it is
$ \Delta \lambda \sim {\rho^2 \over R^3}\sim 20\, MeV $
where the last value correspond to the  mean instanton size $\rho\sim 1/3 \, fm$ and
inter-instanton distance $R\sim 1 \, fm$ \cite{Shuryak:1981ff}.  This fact is especially important for the validity of chiral perturbation theory,
valid only if quark masses are small compared to $ \Delta \lambda$.
 
     This 30-year old (but still little known)  observations were supported by numerical studies of instanton ensembles in 1990's, see review  \cite{Schafer:1996wv}. Lattice observation of the ZMZ states and
     pions  ``made of them" followed a bit later and continue till this day \cite{Ivanenko:1997nb} .  So, topological solitons and  the ZMZ states they generate
are truly responsible for chiral symmetry breaking in QCD.
 The Graz group \cite{Glozman:2012fj}
tried to enforce chiral symmetry by
$removing$ a small strip of near-zero modes from the propagators, making distribution  in Fig.\ref{fig_ZMZ} (left) into
 Fig.\ref{fig_ZMZ} (right). 
Although only a tiny fraction\footnote{It was observed that a handful of those near-zero states
 are also responsible for most of the statistical fluctuations  of (quite expensive) lattice simulation with dynamical fermions.  Therefore creating  more effective theory/algorithms for them would  save millions.
}
 $\sim 10^{-4}$ of all quark states were affected, drastic changes in
  hadronic spectroscopy were observed, with the chiral pairs ($A_1,\rho$), ($N^*(1/2^-), N$)
getting  near-degenerate.    When the inverse thing is done --
 only the contribution {\em  from the strip}  kept in the  propagators --
Graz group found that   $\rho$ and $N$  bound states remained there, albeit with a bit reduced mass.
The latter directly confirms results from the instanton liquid simulations done 20 years ago.

\begin{figure}[!t]\vskip 0.02in
 \center{ \hskip 0in\includegraphics[width=8cm]{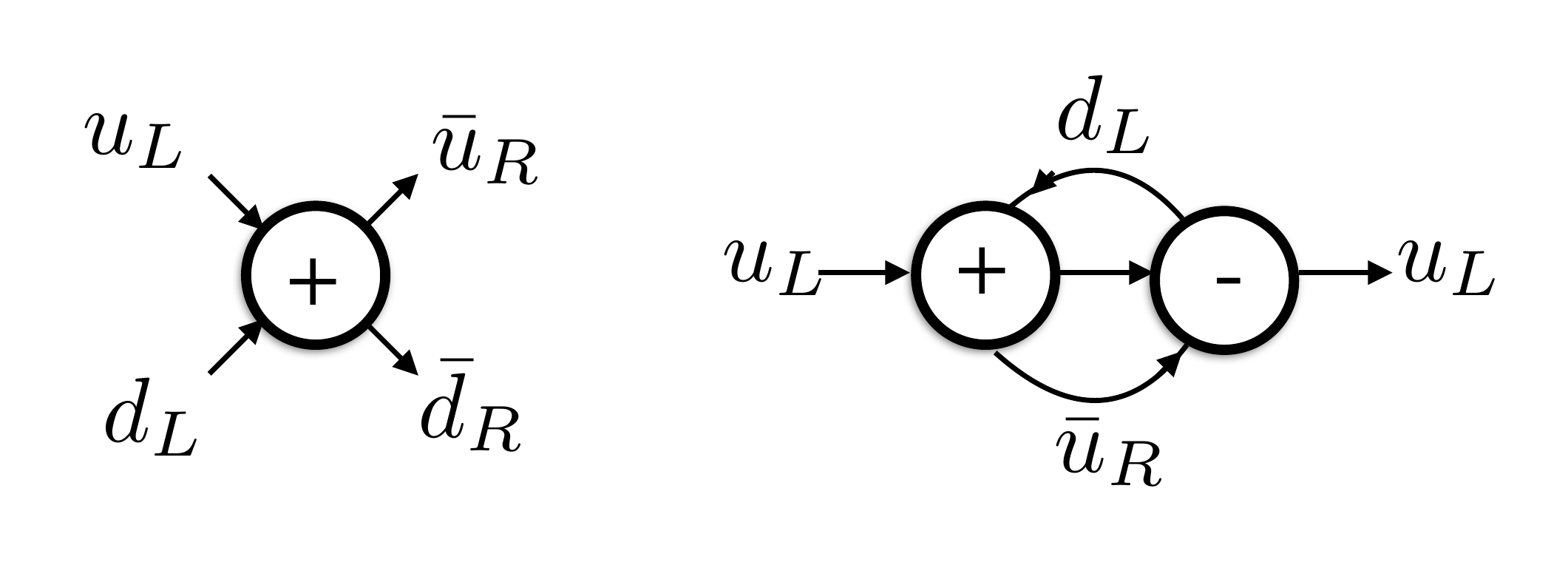}
 }
 \caption{\label{fig_f1} (left) The instanton-induced t' Hooft effective interaction, for $N_f=2$. (right) The effective quark mass
 generated by instanton-antiinstanton molecule. }
 \end{figure}

\subsection{The Origin of Mass}
An "effective mass" of a quasiparticle is usually defined as 
the end of the dispersion curve $M=\omega(k\rightarrow 0)$.  
Delaying discussion of confinement for later, let us imagine we do the same for
light quarks (leading to masses of hadrons made of them). 
 
\begin{figure}[!t]\vskip 0.02in
 \center{ \hskip 0in\includegraphics[width=10cm]{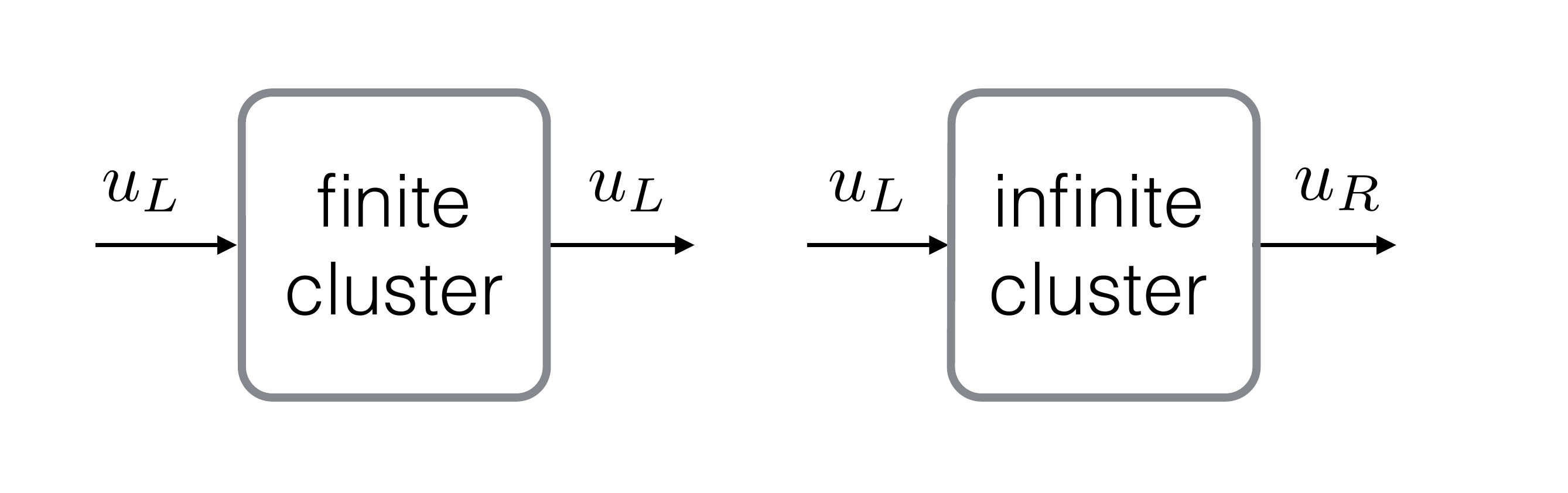}
 }
 \caption{\label{fig_f2} Two types of quark masses: the chirally-even $m_\chi$ (left) and the chirally-odd $m_{\snchi}$ (right). }
 \end{figure}

 A hypothetical 
 4-fermion interaction introduced by Nambu-Jona-Lasinio in 1961 has been identified with the
     instanton-induced   't Hooft  interactions 20 years later  \cite{Shuryak:1981ff}. 
  For a single quark flavor $N_f=1$ (no d) it is produces a mass.    For $N_f=2$
 it is indeed a 4-fermion vertex, as shown in Fig.\ref{fig_f1} (left).
 However, it is not easy to generate a quark mass.
  In the chiral limit, left $L$ and right $R$-handed quarks do not interact
 directly, so one cannot make $\bar{d}d$ into a loop and generate a chirally-odd mass term $\bar{u}_R u_L$.   Adding an
 anti-instanton (as in Fig.\ref{fig_f1}, right) allows to loop unneeded quarks, but  the resulting ``mass operator" -- to which we will return below -- is  in fact chirally-even   $\bar{u}_L u_L$.
Furthermore, {\em any finite number of instantons and antinstantons} generates such terms only. Indeed, one needs the so called {\em infinite cluster} (scaling with volume)  of instantons to
 break  spontaneously the $SU(N_f)$ chiral symmetry. Two types of masses are schematically indicated in Fig.\ref{fig_f2}.

Let me add a couple of digressions before proceeding. The t' Hooft vertex shown in Fig.\ref{fig_f1} (left) 
is also strong attraction in ud scalar diquark channel. This is responsible for
most of the nucleon binding (about 300 MeV) and its quark-diquark structure, as well as
robust color superconductivity. Another comment is that 
 as $N_f$ grows, there are more legs and instanton finite clusters gets
 bound better, As a result, infinite cluster and chiral symmetry breaking disappear. Lattice
studies indeed have difficulty finding any
trace of it for $N_f\sim 8$ and more.

The prevailing philosophy  in  1990's was that the  ``constituent quark mass"   is  chirally odd  $m_{\snchi}$ only.
 Its consequence -- formulated most clearly in the famous Brown-Rho paper \cite{Brown:1991kk}   -- 
is that when chiral symmetry gets restored the quarks (and thus hadrons) are expected to become lighter as well.
Yet several  stubborn observations disagreed. One was the data on chemical
freezeout in heavy ion collisions: while happening near chiral restoration  $T\approx T_\chi$, 
hadron chemistry works fine {\em without} any shifts in masses.   The second was the NA60 dilepton data,
revealing the $\rho$ peak hardly shifted from its vacuum place.
The third is the discovery of 2 solar mass pulsars, requiring extra repulsion in EOS.
All of those were discussed by others at the meeting. Two more came from the lattice. Higher susceptibilities -- derivatives over the baryon chemical potential -- allows to single out the contribution of baryons, and their analysis
\cite{Liao:2005pa} lead us to a conclusion, that as $T$  grows through $T_c$, the nucleon mass does not go down and even
seem to be $increasing$.  The same slowly growing rho and nucleon masses are seen 
by the Graz group already mentioned \cite{Glozman:2012fj}: as  more and more near-zero Dirac eigenstates are removed,  eventually enforcing chiral symmetry to become unbroken.  

So, what is going on here? 
I think we need to include the   chirally-even mass $m_\chi$, coming from the quark energy\footnote{Without the derivative one would get a vector current, 
which is C-odd and can only appear at nonzero baryon density -- e.g. nuclear matter. }
 $m_\chi\sim \bar{q} \partial_0 \gamma_0 q$ which obtains contributions from the finite topological clusters. Near $T_c$, 
of the quark condensate (and infinite cluster) disappears, basically by breaking into finite ones.  Respectively
the non-chiral mass $m_{\snchi}$ partially morphs into the chiral one $m_\chi$, while the total
$M=\sqrt{m_{\snchi}^2+ m_\chi^2}$
perhaps less affected. To calculate all of it, one however needs to do a lot of work, as we discuss below.

\subsection{Nonzero Holonomy, Quasiconfinement  and Instanton-dyons} 
Now we  jump to $T>T_c$ and recall few
important issues. The vacuum average
 of the Polyakov line $P(T)=(1/N_c) Tr <Pexp(i\int A_0^3\tau_3/2 dx^0)>$ changes from 1 to 0 as $T$
decreases through $T_c$ region. 
The ``nonzero holonomy" field\footnote{Here and below we for simplicity discuss only they simplest two color $N_c=2$ gauge theory.} $<A_0>$ changes from zero to 
``confining value"\footnote{For clarity: it is $not$
  the ``chiral mass" we speak about, but an imaginary chemical potential, a phase. 
}
 $\pi T$ at $T_c$.
Furthermore, models including $<A_0>\neq 0$ such as the PNJL model possess $quasiconfinement$:
eliminating quark states from the partition function at  $T<T_c$.  While unable yet
to explain the flux tubes or kill all the gluons, those models clearly are important steps
toward understanding confinement.
 
In 1998 it has been discovered \cite{Lee:1998bb} that a non-vanishing $A_0$
has drastic effect on the gauge topology:  instantons  split into $N_c$ ``instanton dyons",
topological solitons possessing not only topological charge, but also electric and magnetic ones.
(For a review  see \cite{Diakonov:2009ln}.) So, the ``instanton liquid" should evolve into a ``dyonic plasma",
in which instantons -- neutral clusters -- coexist with their ``ionized" constituents.

 Even at a qualitative level, discussed in our first paper  of this program \cite{Shuryak:2012aa}
 the dyons had explained about a dozen puzzling
 lattice observations.  An example: already in 1990's it was observed
 that ``quenched" (no quark determinant) gauge ensembles provide chiral symmetry breaking
at $T<T_{deconfinement}$ for antiperiodic fermions, but not for periodic ones. 
The answer: antiperiodic fermions have zero modes with ``twisted" L dyons, and periodic with ``untwisted" M dyons. Their masses are not the same, $L$ are  heavier  than $M$.
So the same gauge configurations contained two quite different densities of
relevant topological objects, and the denser one keeps chiral symmetry broken to higher $T$.
Another significant benefit is related to the fact that when an instanton measure morphs into a product
of $N_c$ ones for dyons, the transition to large $N_c$ becomes much less singular and puzzling.

The gauge interactions between the instanton-dyons has been discussed in \cite{Diakonov:2009ln} and earlier
papers of Diakonov et al.
Together with Coulombic electric and magnetic forces, there are also confinement-like linear potential induced by the Debye screening by
ambient thermal quarks and gluons. 
The
fermionic exchanges -- related to zero modes -- were worked out in    \cite{Shuryak:2012aa}.  The partition function was finally formulated and the 
first  direct numerical  simulations of the instanton-dyon ensemble were performed
 \cite{Faccioli:2013ja} .  The Dirac eigenvalue spectra were calculated and the necessary conditions -- basically the dyon density -- for the chiral symmetry breaking were identified. 
Finally, let me mention that -- unlike instantons - the instanton-dyons interact directly with
the holonomy field $A_0$. Furthermore, as discussed\footnote{
The idea originated from the paper \cite{Poppitz:2012sw}, in which 
adjoint periodic fermions are used (instead of the fundamental antiperiodic quarks of QCD),
in attempt to build the bridge toward (much simpler) supersymmetric world.
} in \cite{Shuryak:2013tka},
the effective dyon-antidyon repulsion (due to perturbative subtraction) 
has the sign and the magnitude capable to explain why the holonomy shifts from zero
to confinement value, as the dyon density grows.

For the first time we now see how {\em the same}  topological objects are
responsible both for chiral symmetry breaking and  (quasi)confinement. We now see why both phenomena require high enough topological density, albeit maybe different ones at large $N_f$. 
 We are now working on more quantitative derivation,
of an effective PNJL-type model, in which both inputs -- the holonomy potental and the NJL 
term -- are generated dynamically by the instanton-dyons. 

Needless to say, there is a lot to be done.
The instanton-dyons were identified on the lattice, see e.g. \cite{Bornyakov:2013iva} and references therein, but
the information about their density at various temperatures remains sketchy. Certain predictions
made in \cite{Shuryak:2013tka} need to be checked. Intriguing 
generalizations to more fermion types, different fermion boundary conditions or quantum numbers  
can be worked out rather straightforwardly.

\end{document}